\documentclass[review]{elsarticle}
\usepackage{graphicx}
\usepackage{amsmath}
\usepackage{textcomp}
\usepackage{mathtools}
\usepackage{lscape}
\usepackage{subfig}
\usepackage{makecell}
\usepackage{comment}
\usepackage{array}
\usepackage{dirtytalk}
\usepackage{tikz,siunitx}
\usepackage{multirow}
\usepackage{xcolor}
\newcolumntype{P}[1]{>{\centering\arraybackslash}p{#1}}
\newcolumntype{M}[1]{>{\centering\arraybackslash}m{#1}}
\usepackage{framed} 
\usepackage{float}

\usepackage{multicol} 

\usepackage{nomencl} 

\makenomenclature

\renewcommand*\nompreamble{\begin{multicols}{2}}

\renewcommand*\nompostamble{\end{multicols}}
\journal{International Journal of Heat and Fluid Flow}

\begin{document}

\begin{frontmatter}
\title{Compressibility Effects on the Linear-stability of Centrifugal Buoyancy-induced Flow}
\author{Deepak Saini}
\cortext[mycorrespondingauthor]{Corresponding author}
\ead{dsaini@student.unimelb.edu.au}
\author{Richard D. Sandberg}
\address{Department of Mechanical Engineering,\\
The University of Melbourne, Victoria - 3010, Australia}

\begin{abstract}
The focus of this study is to understand the evolution of instability in centrifugal buoyancy-induced flow in a rotating system. The problem is of interest in atmospheric flows as well as in engineering applications. In this study, we perform direct numerical simulations (DNS) by solving the compressible Navier--Stokes equations and multi-dimensional stability analyses by using a forced DNS approach. We systematically and independently vary the Rayleigh and Mach numbers.  The heat transfer by thermal conduction is used as base flow and maintained as a reference state, upon which the growth of small perturbations is investigated. It is found that the critical wavenumber obtained from the linear stability analysis at the onset of convection has a much shorter wavelength than the one that eventually appears in the non-linear regime. Further, the investigations show that compressibility effects lead to a reduction of the growth rate of the dominant mode, and it modifies the overall formation of convection cells in the cavity.

\end{abstract}

\end{frontmatter}
\pagebreak

\section*{Nomenclature} 
\begin{tabbing}
\hspace{1.1in} \= \hspace{0.75in}  \kill
$c_p$ \> specific heat at constant pressure \\
$E$ \> total energy\\
$H$ \> axial gap between the disks \\
$k$ \> wavenumber \\
$L_{\infty}$ \> reference length scale\\
$\lambda$ \> growth rate \\
$m$ \> azimuhtal mode \\
$\Omega$ \> angular velocity\\
$p$ \> pressure\\
$q_{k}$ \> heat flux vector \\
$r$ \> radius or radial direction\\
$\rho$ \> density\\
$S$ \> Sutherland temperature \\
$T$ \> temperature \\
$\tau_{ik}$ \> viscous stress tensor \\
$t$ \> non-dimensional time units \\
$u,v,w$ \> velocity in $z$, $r$ and $\theta$ direction \\
$U_{\infty}$ \> free fall velocity\\
$\mu$ \> dynamic viscosity\\
$\kappa$ \> thermal conductivity \\
$p$ \> pressure\\
\end{tabbing}
\subsection*{Dimensionless Numbers}
\begin{tabbing}
\hspace{1.1in} \= \hspace{0.75in}  \kill
$\gamma$ \> ratio of specific heats\\
$Ma$ \> Mach number \\
$Nu$ \> Nusselt number\\
$Pr$ \> Prandtl number\\
$Ra$ \> Rayleigh number\\
$Re$ \> Reynolds number\\
$Ro$ \> Rossby number\\
\end{tabbing}

\subsection*{Subscripts and Superscripts }
\begin{tabbing}
\hspace{1.1in} \= \hspace{0.75in}  \kill
$ad$ \> adiabatic \\
$C$ \> at the cold wall\\
$H$ \> at the hot wall\\
$\infty$ \> reference quantity\\
$i$ \> at the inner radius\\
$o$ \> at the outer radius\\
$m$ \> at the mean radius\\
\end{tabbing}
\subsection*{Abbreviation}
\begin{tabbing}
\hspace{1.1in} \= \hspace{0.75in}  \kill
$DNS$ \> Direct Numerical Simulation\\
$LSA$ \> Linear Stability Analysis\\
\end{tabbing}

\section{Introduction}
\label{intro}
\begin{figure}[H]
\vspace{2mm}
\centering
\includegraphics[width=0.7\textwidth]{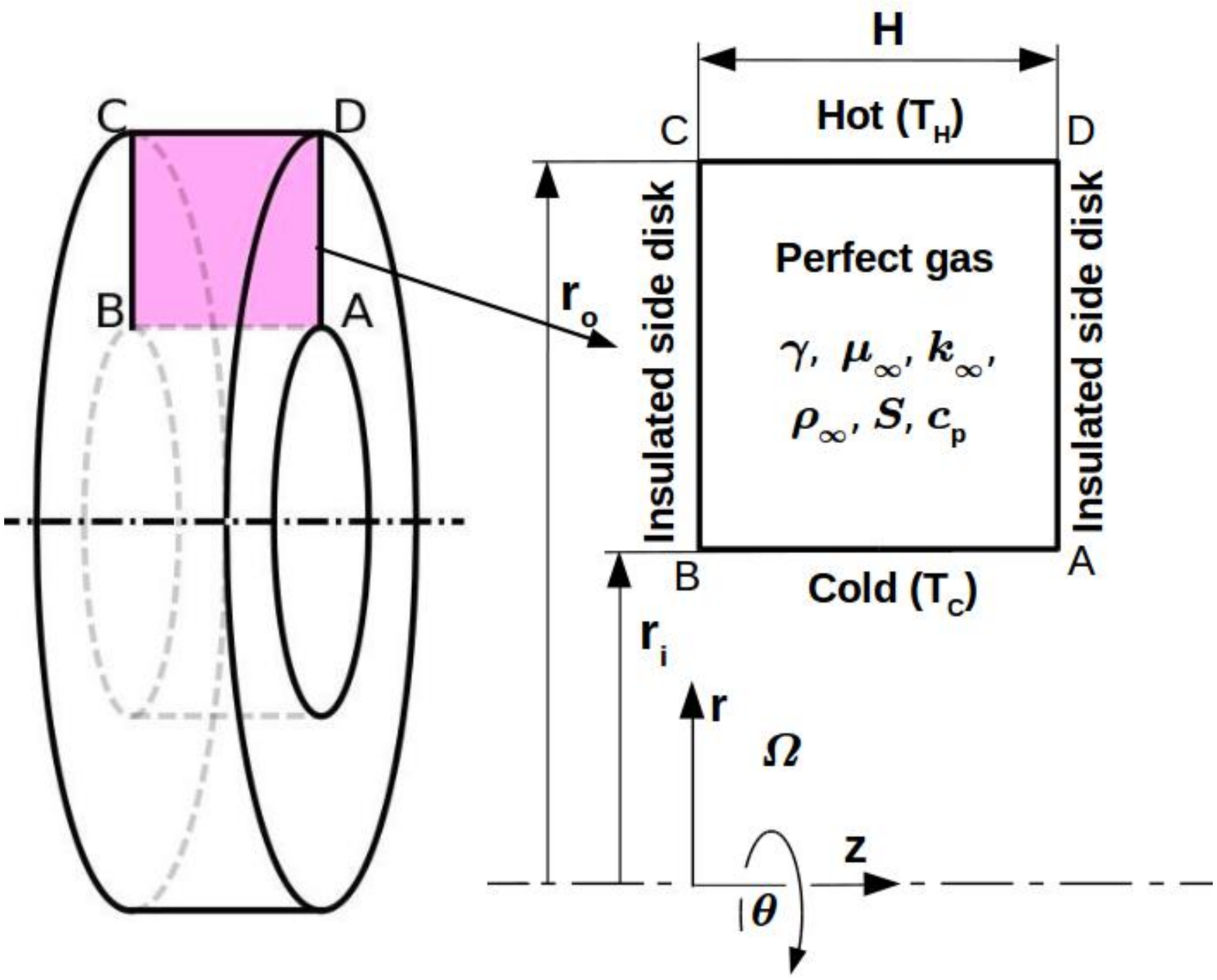}
\vspace*{-0.1in}
\caption{Schematic of closed annular cavity. The walls of the cavity undergo solid-body rotation about the $z$-axis.}
\label{fig:sch}
\end{figure}

Centrifugal buoyancy induced convection is a prevalent phenomenon in geophysical fluid dynamics as well as in engineering applications. It controls numerous physical phenomena such as geomagnetism \cite{busse2002convective} and deep convection \cite{jonathan2008convective} in planetary flows and is also crucial for the understanding the global circulations in the atmosphere \cite{read2015general}. 
The flow-induced by centrifugal buoyancy is also cardinal across a wide range of engineering applications, such as gas turbines \cite{owen2015review}, and it directly influences critical performance parameters of various engineering products. The governing parameters in such kind of flows, such as imposed temperature gradient and rotation rate beyond their critical values, lead to a flow instability that grows to form convection cells. In this study, we adopt a geometrical configuration, as shown in figure \ref{fig:sch}. The outer cylindrical wall is maintained at a high temperature, and the inner cylindrical wall is at a low temperature, bounded by two insulated side disks. The geometrical configuration is primarily motivated by the rotor/rotor cavities of the internal air system \cite{owen2015review} of a gas turbine.

Initial experiments for the centrifugal buoyancy-induced flows were carried out by Busse and Carrigan \cite{busse1974convection}. They studied the onset of convection in a rotating annulus with a heated outer wall and a cold inner wall. They proposed an analytical expression based on the linear theory for the critical value of the Buoyancy parameter as a function of Ekman number and Prandtl number. The agreement between the experimental observations and the analytical results suggested that linear theory is able to describe such kinds of flows.
Eckhoff and Storesletten \cite{eckhoff1980stability} investigated the flow stability for an inviscid rotating fluid to understand the effects of compressibility. They proposed a stability condition applicable to rotating systems in a gravitational force field such as planetary and stellar atmospheres. Most of the subsequent studies were mainly motivated by geophysical flows \cite{read2015general} and in such conditions, both centrifugal force and gravity were considered to underpin the flow dynamics. However, we are particularly interested in the conditions where the magnitude of centrifugal forces is much larger than the gravitational force such that we can neglect the latter. Under these conditions, the compressibility effects in the direction of centrifugal force become significant when rotational speed increases to a value beyond a critical number.

An initial investigation on the effect of high rotational speed in a rotating cavity was highlighted by Kilfoil and Chew \cite{kilfoil2009modelling}. They proposed a stability condition for centrifugal-buoyancy induced flow and then used it to accurately simulate such flows by adding an additional term in the momentum equation. Another numerical study for centrifugal buoyancy-induced flows was carried out by Pitz et al. \cite{pitz2017onset} for the onset of convection. They performed linear stability analysis based on the incompressible Navier--Stokes equations to obtain the critical Rayleigh number at different radius ratios. In the asymptotic limit of radius ratio approaching unity, the critical Rayleigh number was observed to be $1708$. This value also corresponds to the critical value for natural convection under gravity. Further, the critical wavenumber number obtained from the linear stability analysis was observed to be dominant in the turbulent regime as well. 

Another study that carried out linear stability analyses by solving incompressible equations with the Boussinesq approximation is by Kang et al. \cite{kang2019numerical}. They included an additional Coriolis buoyancy term in their equations and investigated the annular cavity problem for a Rayleigh number of the order of $10^5$. A decrease in the value of the critical Rayleigh number with the increase in the radius ratio was observed. They also found that the critical Rayleigh number approached the value of $1708$ in the limit of radius ratio approaching unity. A complex transition phenomenon from the onset of the convection to the chaotic motion was reported.  

The studies mentioned above with an emphasis on the linear stability were limited to the physical conditions where the Boussinesq approximation was valid and thus considered Rayleigh number as the only critical parameter. 
However, at high rotational speeds, the effects of compressibility play a significant role and affect the flow structure and heat transfer rate.
The experimental evidence for this in the case of open cavities, which also have direct relevance to internal air system of a gas turbine engine, was reported by Farthing et al. \cite{farthing1992rotating_HT} for engine representative conditions and has been recently observed by Jackson et al. \cite{jackson2020measurement} in their experimental measurements. They observed a decrease in the value of wall Nusselt number with an increase in the value of the rotational Reynolds number. They attributed this to the compressibility effects in the core of the cavity, as it led to an increase in the core temperature and reduced the wall heat transfer rate. Tang and Owen \cite{tang2018theoretical} in their theoretical model considered the compressibility effects in the core to predict the wall Nusselt number. They proposed a scaling of the Nusselt number as
\begin{equation}
Nu \propto Ra^{1/4} (1-\zeta Re^2)^{5/4},
\label{tang}
\end{equation}
where $\zeta$ was the compressibility parameter. They concluded that the compressibility has a direct influence on the wall heat transfer rate as given by equation \ref{tang}. However, they sought more experimental or computational evidence to support their hypothesis about the effects of compressibility in centrifugal buoyancy-induced flows. Using direct numerical simulation, Saini and Sandberg \cite{saini2020simulations} further analyzed the compressibility effects. They reported that the radial density distribution approached stratified conditions with an increase in the flow Mach number, and this prevented the convection rollers from growing. They concluded that there is a direct influence of compressibility effects on the growth of the flow instability.  A very recent numerical study by Gao and Chew \cite{gao2021ekman} also highlighted the compressibility effects in a closed cavity. The authors conjectured that at high $Ra$ number, the compressibility effects reduced the driving temperature gradient for convection and reduced the likelihood of flow transition to a fully turbulent state.

Clearly, most of the studies on the linear stability of centrifugal buoyancy-induced flow were limited to incompressible flow. Few studies highlighted the role of compressibility in the evolution of hydrodynamic instability, but those studies lacked a detailed investigation. 
Hence, this study aims to understand the effects of compressibility on the linear growth of flow instability in the centrifugal buoyancy-induced flow by solving the compressible Navier--Stokes equations and systematically varying Mach number for different Rayleigh numbers. .

\section{Problem description and Numerical setup}
The computational domain in this study is a closed annular cavity (figure \ref{fig:sch}) that rotates at a constant angular velocity of $\Omega$ about the z-axis. The outer wall at radius $r_o$ of the cavity is at temperature $T_{H}$ and the inner wall at radius $r_i$ is at temperature $T_{C}$. The temperature difference between the outer and the inner wall ($T_{H}-T_{C}$), and the centrifugal acceleration $\Omega^2 r$, set up the centrifugal buoyancy-induced flow in the cavity. To study the onset of convection, we carried out linear stability analysis. The approach followed in this study is discussed in the next section.

\subsection{Linear stability analysis} 
The multi--dimensional stability analyses are carried out by using a forced Navier--Stokes simulation approach. We consider a base state of solid body rotation of the cavity with conduction heat transfer between the outer and the inner wall. At time $t=0$, the base state for the conservative variables $\rho^{0}u_{i}^{0}$, $\rho^{0}E^{0}$ and density $\rho^{0}$ is
\begin{equation}
\begin{aligned}
&\ \rho^{0} = 1.0,\quad \rho^0 u_{1}^{0} = 0.0, \quad \rho^0 u_{2}^{0} = 0.0, \quad \rho^0
u_{3}^{0} = \Omega r, \\
&\ \rho^{0} E^{0} = \frac{\rho^{0} T^{0}}{\gamma(\gamma-1) Ma_{\infty}^2} + (1/2)\rho^{0} u_{i}^{0} u_{i}^{0},
\end{aligned}
\label{BS}
\end{equation}
where $T^{0} = (T_{H}-T_{C}) [log(r/r_o)/log(r_o/r_i)]+T_{H}$ is the heat conduction temperature profile.
At the beginning of the simulation, all the spatial derivatives of the base state for the continuity, momentum and energy equations are calculated as
\begin{equation}
\frac{\partial \rho}{\partial t} \Bigr\rvert_{t = 0}  = - \frac{\partial}{\partial x_{k}} (\rho^{0} u_{k}^{0}),
\label{1t0}
\end{equation}
\begin{equation}
\frac{\partial}{\partial t}(\rho u_{i})\Bigr\rvert_{t = 0} =  - \frac{\partial}{\partial x_{k}}[\rho^{0} u_{i}^{0} u_{k}^{0} + p^{0} \delta_{ik} - \tau_{ik}^{0}],
\label{2t0}
\end{equation}
\begin{equation}
\frac{\partial}{\partial t}(\rho E) \Bigr\rvert_{t = 0} = - \frac{\partial}{\partial x_{k}}[u_{k}^{0} (\rho^{0} E^{0}+p^{0}) + q_{k}^{0} - u_{i}^{0}\tau_{ik}^{0}],
\label{3t0}
\end{equation}
and stored. The viscous stress tensor and the heat-flux vector are computed as
\begin{equation}
\begin{aligned}
&\ \tau_{ik} = \frac{\mu}{(Ra_{\infty}/Pr_{\infty})^{1/2}}\bigg(\frac{\partial u_{i}}{\partial x_{k}} + \frac{\partial u_{k}}{\partial x_{i}} - \frac{2}{3} \frac{\partial u_{j}}{\partial x_{j}}\delta_{ik}\bigg), \\
&\  q_{k}= \frac{-\mu}{(\gamma-1)Ma_{\infty}^{2} (Ra_{\infty} Pr_{\infty})^{1/2}} \frac{\partial T}{\partial x_{k}}.
\end{aligned}
\label{4}
\end{equation}
In the above equations, $Ra_{\infty}$, $Pr_{\infty}$ and $Ma_{\infty}$ are the reference Rayleigh, Prandtl and Mach numbers, respectively. The definitions of the above-mentioned non-dimensional numbers and other parameters are given as
\begin{equation}
\begin{aligned}
&\ Ra_{\infty} = \frac{\rho_{\infty}^{2} c_p (\Omega^2 r_m)(\Delta T/T_{\infty})(r_o-r_i)^3}{\mu_{\infty} \kappa_{\infty}},\quad \gamma = 1.4, \\ 
&\ Pr_{\infty} \equiv \frac{\mu_{\infty} c_{p}}{\kappa_{\infty}}=0.7 ,\quad \frac{\Delta T}{T_{\infty}},\quad Ma_{\infty} \equiv \frac{U_{\infty}}{(c_{p}(\gamma-1)T_{\infty})^{1/2}}.
\end{aligned}
\label{ND}
\end{equation}
The effect of Coriolis forces in the system is characterized by the Rossby number defined as
\begin{equation}
\begin{aligned}
&\ Ro \equiv \frac{[\Omega^2 r_m (\Delta T/T_{\infty})(r_o-r_i)]^{1/2}}{2 \Omega (r_o - r_i)}.
\end{aligned}
\label{Ro_eq}
\end{equation}

In equation \ref{ND}, $L_{\infty}=r_{o}-r_{i}$, $U_{\infty}=\big[\Omega^2 r_m (\Delta T/T_{\infty})L_{\infty}\big]^{1/2}$ and $T_{\infty}=(T_{H}+T_{C})/2$ are the reference length, velocity, and temperature scales, respectively. Further, $r_i/L_{\infty}=1.0869$, $r_o/L_{\infty}=2.0869$ and $H/L_{\infty}=1.043$ are the non-dimensional inner radius, outer radius and the axial length of the cavity.  The properties of a perfect gas with specific heat at constant pressure $c_{p}$ and the ratio of specific heats $\gamma$, molecular viscosity $\mu_{\infty}$ and the thermal conductivity $\kappa_{\infty}$ are defined at the reference temperature $T_{\infty}$.  
To close the above system of equations, the pressure is computed from the equation of state  $ p = \rho T/ (\gamma Ma_{\infty}^{2})$. The temporal derivatives of the base state defined in equations \ref{1t0}, \ref{2t0} and \ref{3t0} are maintained as a reference state upon which the behaviour of small perturbations is investigated by subtracting it from the solution of the Navier--Stokes equations at each Runge--Kutta substep as the solution progresses in time as follows

\begin{equation}
\frac{\partial \rho}{\partial t} = - \frac{\partial}{\partial x_{k}} (\rho u_{k})-\frac{\partial \rho}{\partial t} \Bigr\rvert_{t = 0},
\label{1t0n}
\end{equation}
\begin{equation}
\frac{\partial}{\partial t}(\rho u_{i}) =  - \frac{\partial}{\partial x_{k}}[\rho u_{i} u_{k} + p \delta_{ik} - \tau_{ik}]-\frac{\partial}{\partial t}(\rho u_{i})\Bigr\rvert_{t = 0},
\label{2t0n}
\end{equation}
\begin{equation}
\frac{\partial}{\partial t}(\rho E) = - \frac{\partial}{\partial x_{k}}[u_{k} (\rho E+p) + q_{k} - u_{i}\tau_{ik}]-\frac{\partial}{\partial t}(\rho E) \Bigr\rvert_{t = 0}.
\label{3t0n}
\end{equation}

The equations are discretised with a fourth-order accurate finite-difference scheme in the axial $(z)$ and radial $(r)$ directions.  A Fourier spectral method is used for discretisation in the azimuthal ($\theta$) direction.
For the stability analysis, all the Fourier modes are perturbed by a spatial step function in the $r$--$z$ plane of amplitude $10^{-6}$ at the center of the cavity. The perturbation is introduced at the initial state and then its development is monitored as the simulation progresses. If the flow conditions are unstable then the initial pertubation would grow exponentially until it reaches a non-linear state. In contrast, for the stable flow conditions the response of an initial perturbation subsides with time until it disappears.
This approach has been successfully used in previous studies to investigate the linear stability of flow over an aerofoil \cite{jones2010stability} and for axisymmetric wakes \cite{sandberg2012numerical}.

\section{Results and Discussion}
We have carried out the linear stability analyses (LSAs) for two $Ra$ numbers of $10^5$ and $10^7$ at a constant $Ro$ number of $0.199$ to understand the onset of convection. To further study the compressibility effects, we varied the $Ma$ number in the range $0.01$--$0.2$ for $Ra=10^5$ and $0.05$--$0.2$ for $Ra=10^7$, respectively. The grid resolution in the $r-z$ directions is kept the same as for the DNS \cite{saini2020simulations}. In the azimuthal direction, $32$ Fourier modes are used for both the $Ra$ numbers. 
\begin{figure}[H]
\centering
\includegraphics[width=0.85\textwidth]{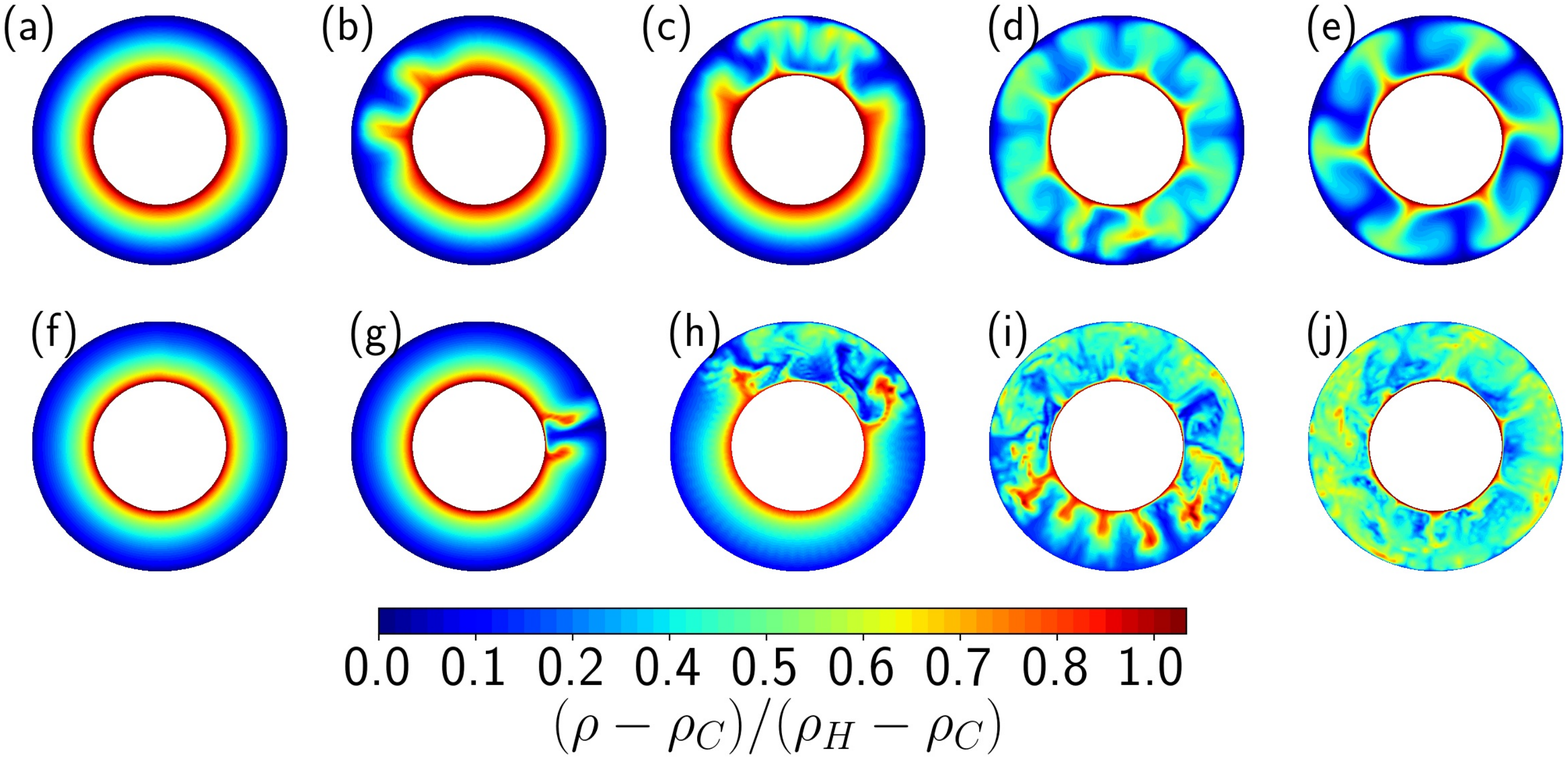}
\includegraphics[width=0.75\textwidth]{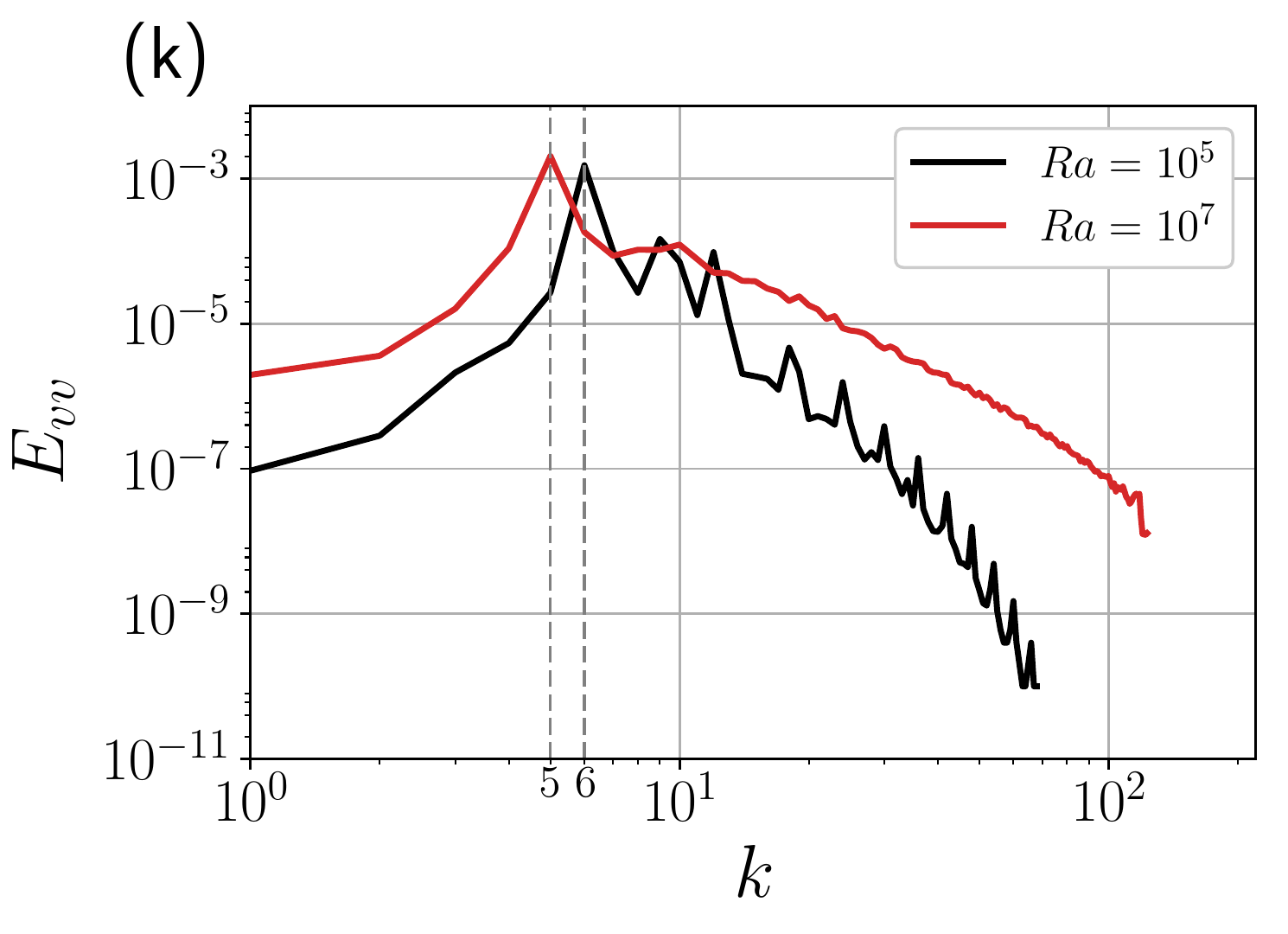}
\vspace*{-0.1in}
\caption{Time evolution of density at the mid-axial position $z = (1/2)(H/L_{\infty})$ of the cavity for $Ra = 10^5$, $Ro=0.199$ and $Ma=0.01$ at non-dimensional time units (a-e) $t=10.0$, $22.0$, $25.0$, $40.0$ and $55.0$, respectively, and for $Ra = 10^7$, $Ro=0.199$ and $Ma=0.05$ at non-dimensional time units (f-j) $t=12.5$, $18.0$, $25.0$, $32.5$ and $38.0$, respectively. (k) Wavenumber spectra of the monitor point located at the middle of the cavity at different Rayleigh numbers.  All results from DNS \cite{saini2020simulations}, which has higher number of Fourier modes in the azimuhtal direction.}
\label{fig:T_eve_05}
\end{figure}

\begin{figure}[ht]
\centering
\includegraphics[width=0.85\textwidth]{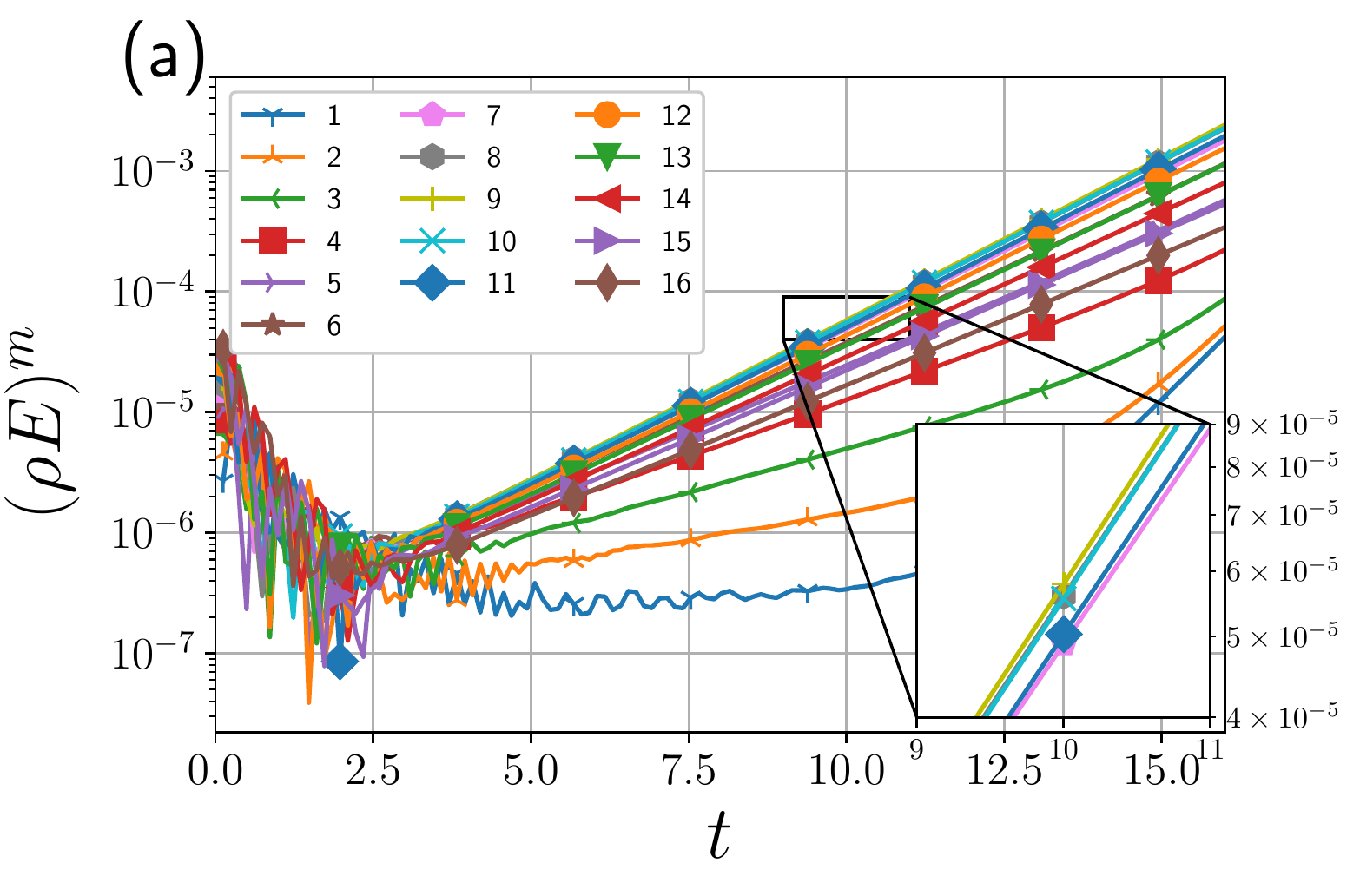}
\includegraphics[width=0.85\textwidth]{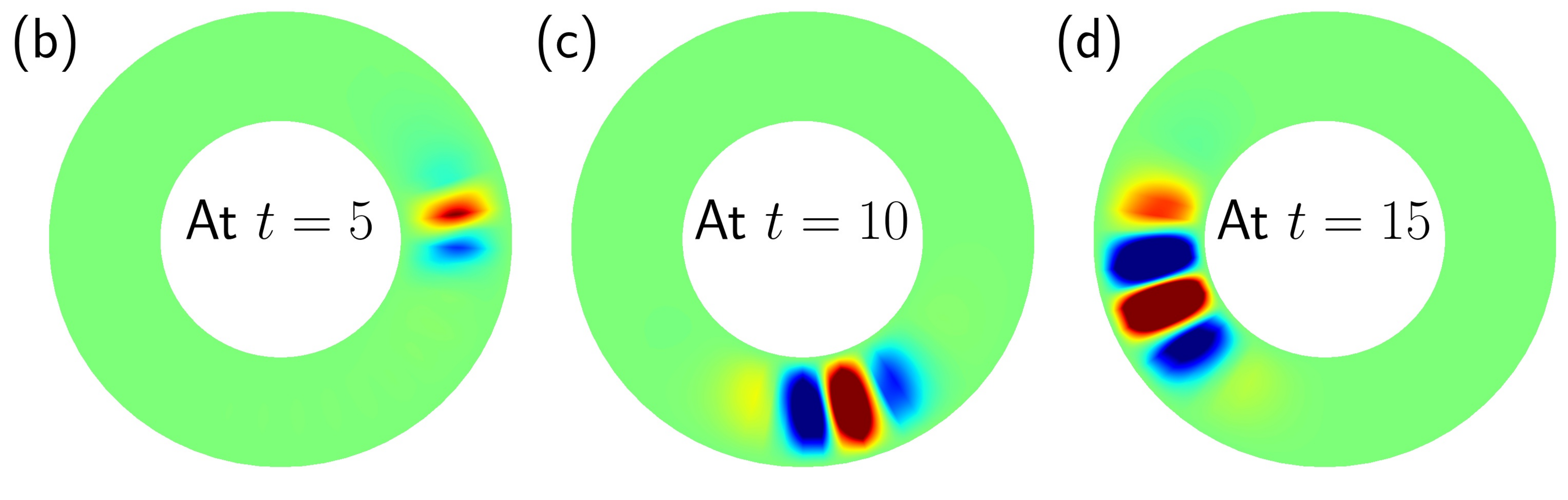}
\vspace*{-0.1in}
\caption{(a) Temporal developement of first $16$ most amplified Fourier modes in the LSA at $(z,r)=(0.527,1.2)$ for $Ra=10^5$ at $Ma=0.01$ and $Ro=0.199$. Instantaneous contour for radial velocity fluctuations at (b) $t=5$ (contour levels [$-1.34e-05$, $+1.34e-05$]), (c) $t=10$ (contour levels [$-2.34e-05$, $+2.34e-05$]) and (d) $t=15$ (contour levels [$-9.34e-05$, $+9.34e-05$]). }
\label{fig:E_LSA_05}
\end{figure}
Figure \ref{fig:T_eve_05} shows the time evolution of instability from the DNS for $Ra=10^5$ and $Ra=10^7$ cases. The initial state of heat conduction in the DNS is perturbed by adding sinusoidal perturbations of the form  $sin (k \theta)$ to the temperature. From the instantaneous flow field, the initial conduction state of the system can be observed in figure \ref{fig:T_eve_05}(a). As the flow evolves in time, the local disturbance added to the conduction state at the interface of hot and cold fluid can be seen (figure \ref{fig:T_eve_05}(b)). This localized disturbance or the instability in the flow, which is at a fetal stage at this time, further grows temporally and spatially to form ``mushroom'' shaped structures as shown in the figure \ref{fig:T_eve_05}(c). The plumes of the high-density fluid (low temperature) rise in an asymmetric manner in the radial direction (figure \ref{fig:T_eve_05}(c)). At the fully grown stage (figure \ref{fig:T_eve_05}(d)), there are eight plumes in the domain. Similar behaviour for the time evolution of the instability can be observed in the case of $Ra=10^7$ in figures \ref{fig:T_eve_05}(f)-(j). At higher $Ra$ number, the flow transitions to turbulence more rapidly, and a more chaotic flow field emerges as shown in figure \ref{fig:T_eve_05}(i). At the fully developed state, the number of convection rollers is six for $Ra = 10^5$ and five for $Ra = 10^7$, which also corresponds to the peak energy level in the time-averaged wavenumber spectra at $k=6$ and $k=5$, respectively, as shown in figure \ref{fig:T_eve_05}(k). The multiple peaks observed in the wavenumber spectra for $Ra=10^5$ case correspond to the harmonics of the dominant wavenumber, which can be distinctly observed in the spectra as the flow is not fully turbulent.

\begin{figure}[ht]
\centering
\includegraphics[width=0.85\textwidth]{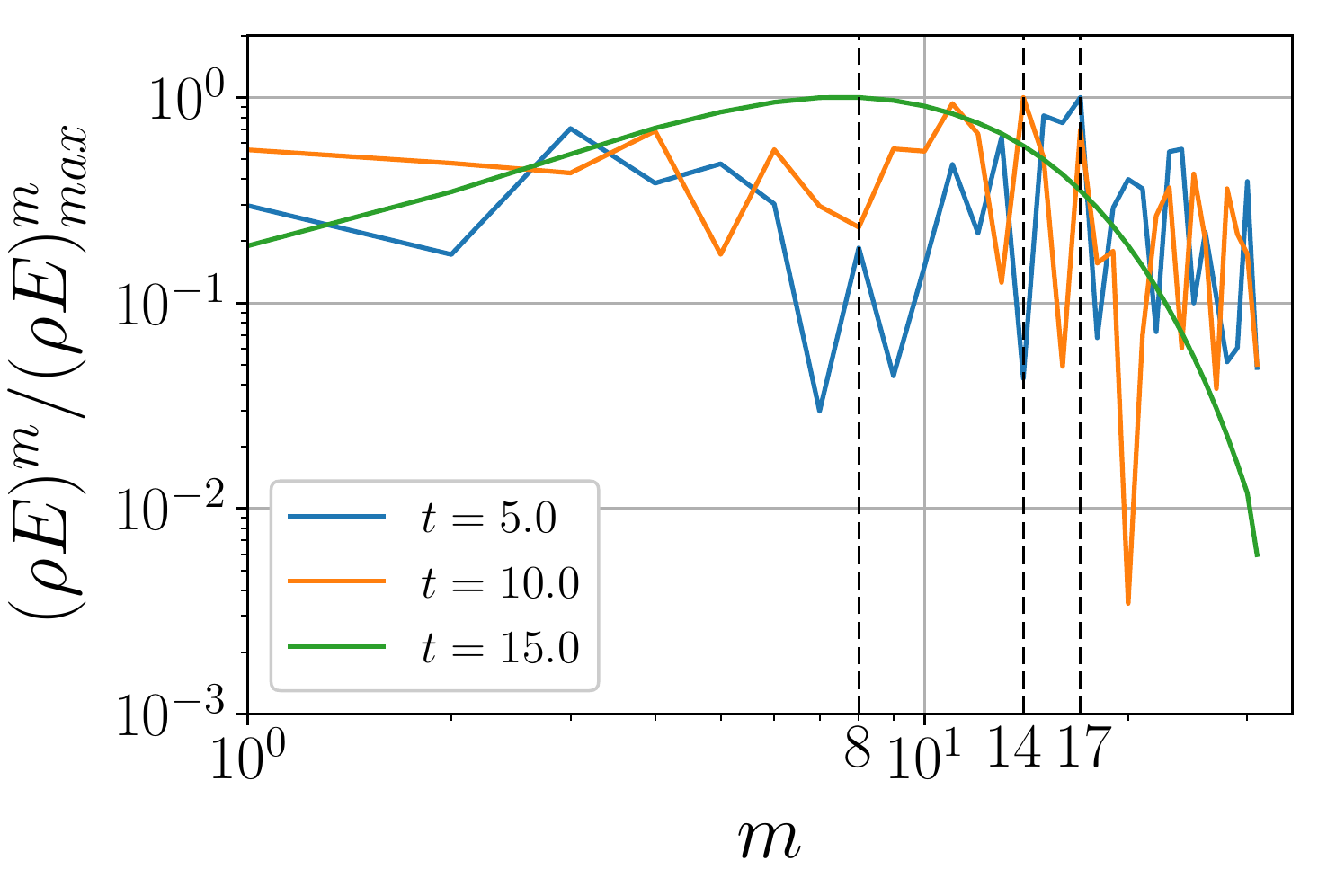}
\vspace*{-0.15in}
\caption{The azimuthal spectra of the monitor point located at the middle of the cavity at different time instants to highlight the distribution of energy in azimuthal Fourier modes for $Ra=10^5$ at $Ma=0.01$ and $Ro=0.199$. }
\label{fig:Spec_LSA_05}
\end{figure}

\begin{figure}[ht]
\centering
\includegraphics[width=0.85\textwidth]{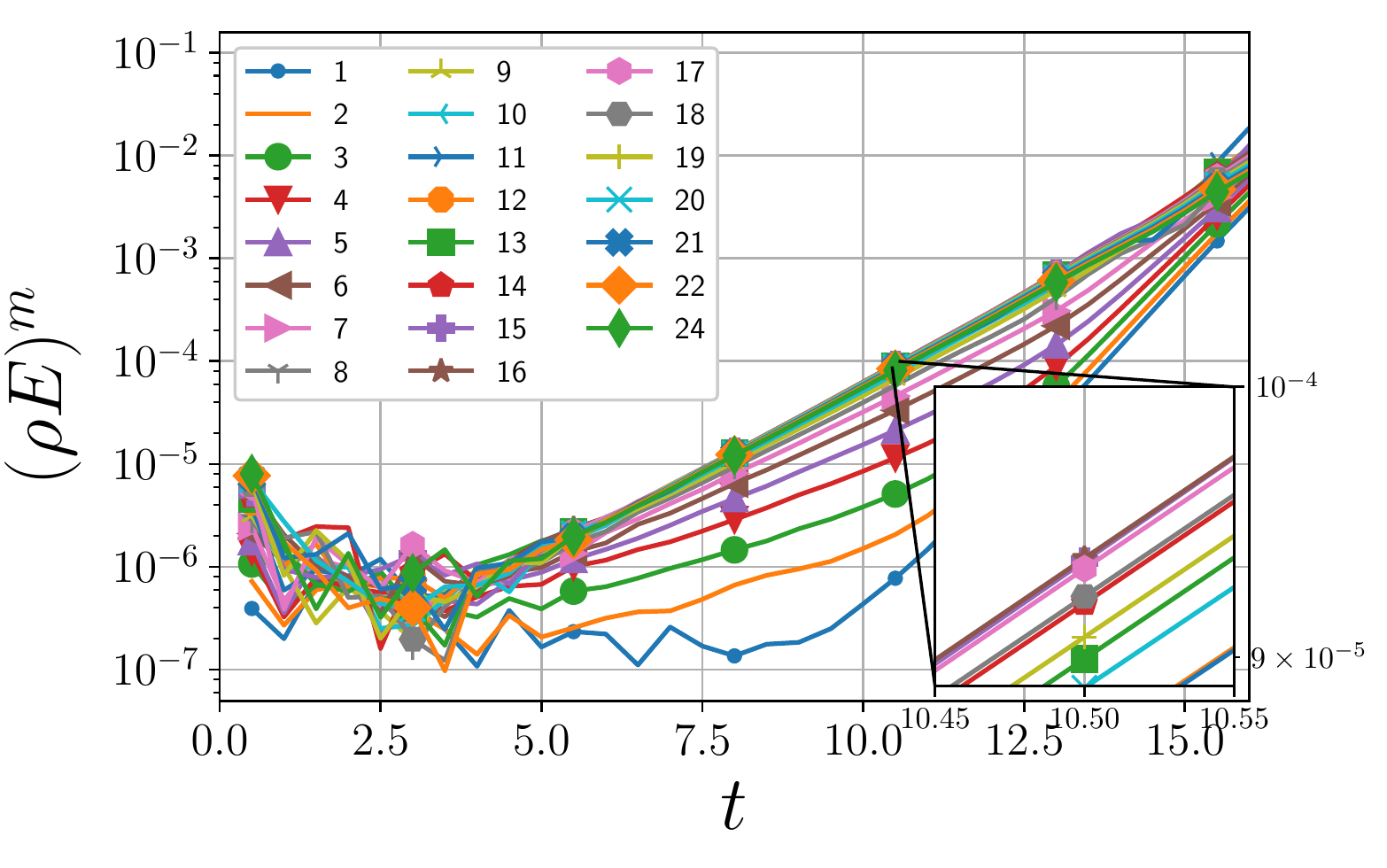}
\vspace*{-0.15in}
\caption{Temporal developement of first $24$ most amplified Fourier modes in the LSA at $(z,r)=(0.527,1.2)$ for $Ra=10^7$ at $Ma=0.05$ and $Ro=0.199$. }
\label{fig:E_LSA_07}
\end{figure}

\begin{figure}[ht]
\centering
\includegraphics[width=0.85\textwidth]{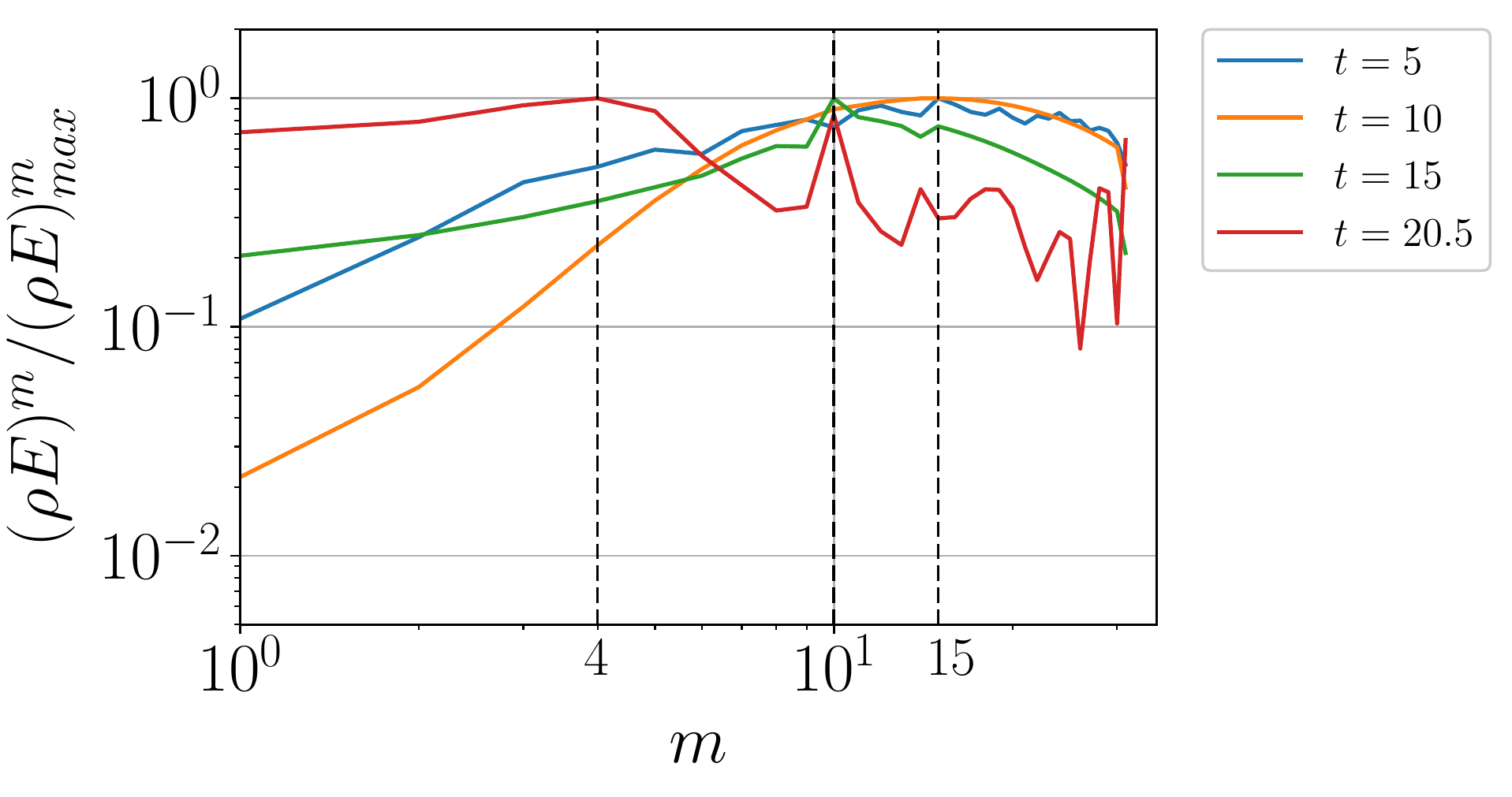}
\vspace*{-0.15in}
\caption{The azimuthal spectra of the monitor point located at the middle of the cavity at different time instants to highlight the distribution of energy in azimuthal Fourier modes for $Ra=10^7$ at $Ma=0.05$ and $Ro=0.199$. }
\label{fig:Spec_LSA_07}
\end{figure}

To understand this transition from the initial conduction state to an unsteady flow field, we carried out linear stability analysis for the same conditions. We perturbed all the modes with an initial pulse of amplitude  $10^{-6}$ at the $r$--$z$ location of $(1.586,0.5217)$. The behaviour of all the azimuthal Fourier modes is then monitored. Figure \ref{fig:E_LSA_05}(a) shows the temporal development of the first sixteen most-amplified azimuthal modes. The wavenumbers $k=8$, $9$ and $10$ are the most unstable and have the highest growth rates. 
Further, as observed in figure \ref{fig:E_LSA_05}(a), multiple modes get amplified in the linear regime, and the superposition of these modes leads to an asymmetric or localized growth of the instability.
The localized spatial growth of the instability can be observed in the figures \ref{fig:E_LSA_05}(b)-(d), which show the instantaneous contours of radial velocity fluctuations at different time instants. The disturbance at $t=5.0$ in the radial velocity is localized, and the energy is concentrated in high wavenumbers.  Further, at $t=10.0$ and $20.0$, the instability grows spatially, as shown in figure \ref{fig:E_LSA_05} (c)-(d), with an increase in the size of rollers and temporal growth of the disturbance can be inferred from the increase in the energy of most-amplified modes in figure \ref{fig:E_LSA_05}(a). The azimuthal spectra at different time instants are plotted in figure \ref{fig:Spec_LSA_05} to understand the energy distribution across different azimuthal modes. At $t=5.0$ and $10.0$, most of the energy is contained in higher modes ($m>10$) with a peak energy level in modes $m=17$ and $14$, respectively. However, as time progresses, a redistribution of the energy takes place, and maximum energy appears at $m=8$ at $t=15.0$. 
Similar behaviour is observed for the $Ra=10^7$ case in figure \ref{fig:E_LSA_07}. In this case, the wavenumbers $k=15$, $16$ and $17$ are the most linearly amplified. Also, in this case, multiple Fourier modes with higher wavenumbers get amplified in the initial phase of flow transition from a stable state to an unstable state. To illutrate this, the spectra of azimuthal modes at different time instants are shown in figure \ref{fig:Spec_LSA_07}. In the initial time instants at $t=5.0$ and $10.0$, higher modes ($m >10$) contain most of the energy with a peak energy level in $m=15$. This peak further shifts towards lower wavenumber at $t=15$ and $20.5$, where the azimuthal modes $m=10$ and $4$, respectively, attain the maximum energy.

Overall, in the linear regime, the most unstable wavenumbers have a much shorter wavelength than the one that eventually appears in the fully developed state in the DNS. This implies that non-linear interaction between these linearly unstable modes occurs, which eventually leads to amplification of the lower wavenumber in the saturation state.

\begin{figure}[ht]
\centering
\includegraphics[width=0.85\textwidth]{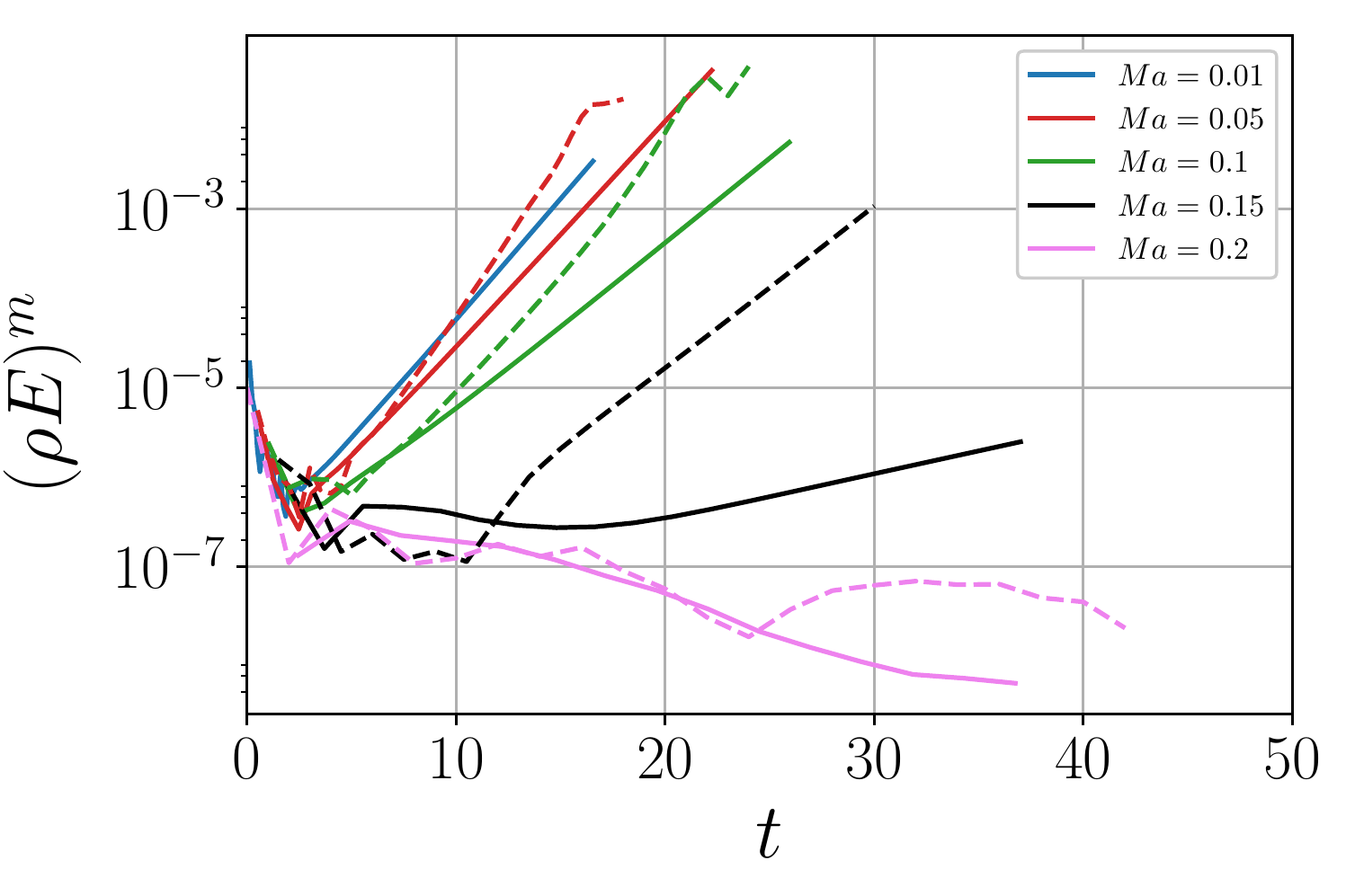}
\vspace*{-0.15in}
\caption{Time evolution of dominant mode in the azimuthal direction at $(z,r)=(0.527,1.2)$ for \protect\tikz[baseline]{\protect\draw[thick] (0,.5ex)--++(.5,0) ;} {$Ra=10^5$},
\protect\tikz[baseline]{\protect\draw[thick][dashed] (0,.5ex)--++(.5,0) ;} {$Ra=10^7$} for $Ro=0.199$ at different $Ma$ numbers.}
\label{fig:Ma1e05}
\end{figure}

\subsection{Effect of compressibility} 
\label{eff_comp}
\begin{figure}[ht]
\centering
\includegraphics[width=0.85\textwidth]{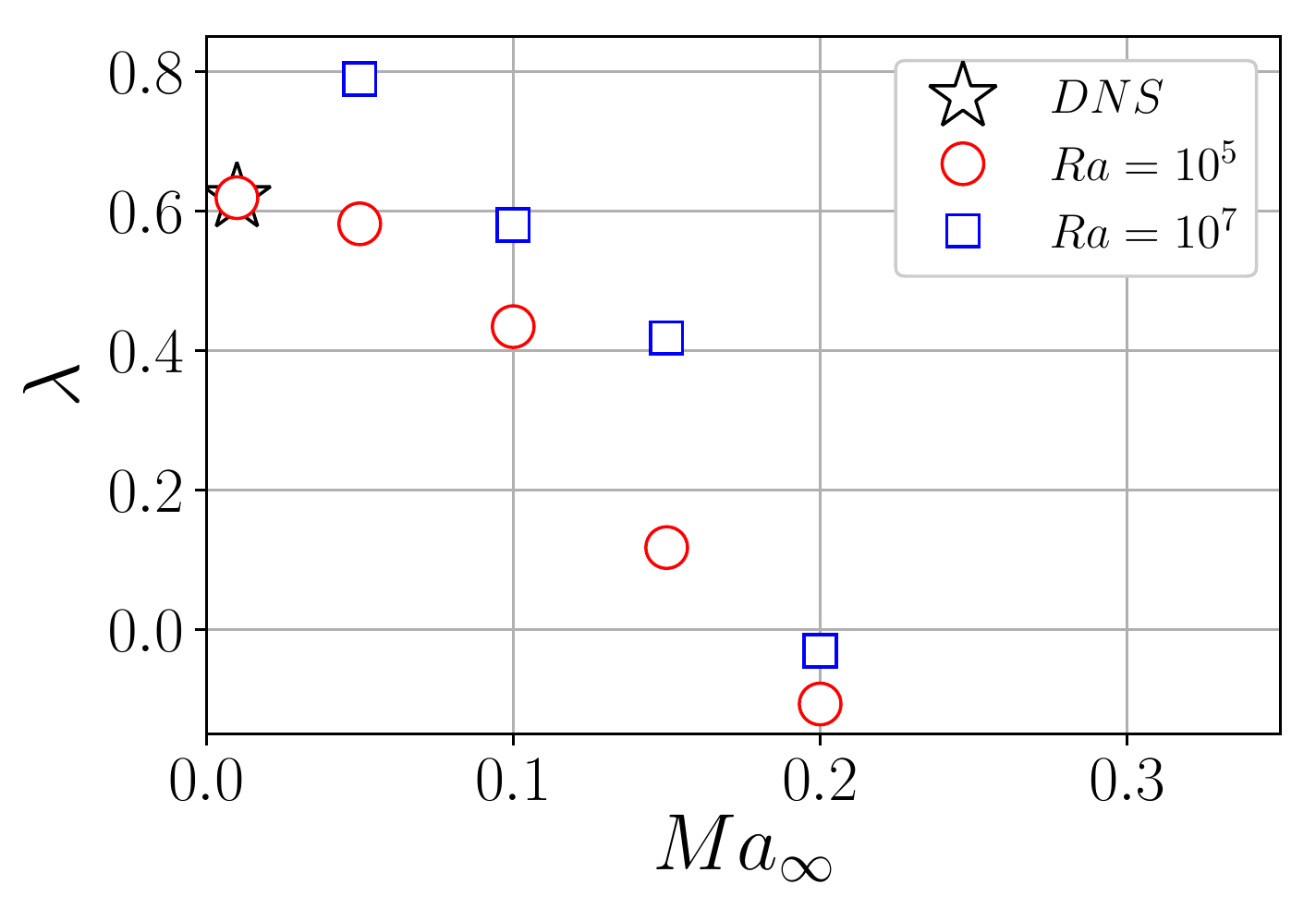}
\vspace*{-0.15in}
\caption{Comparison of growth rate of most unstable mode for different $Ra$ numbers at $Ro=0.199$.}
\label{fig:lambda}
\end{figure}

\begin{figure}[ht]
\centering
\includegraphics[width=0.8\textwidth]{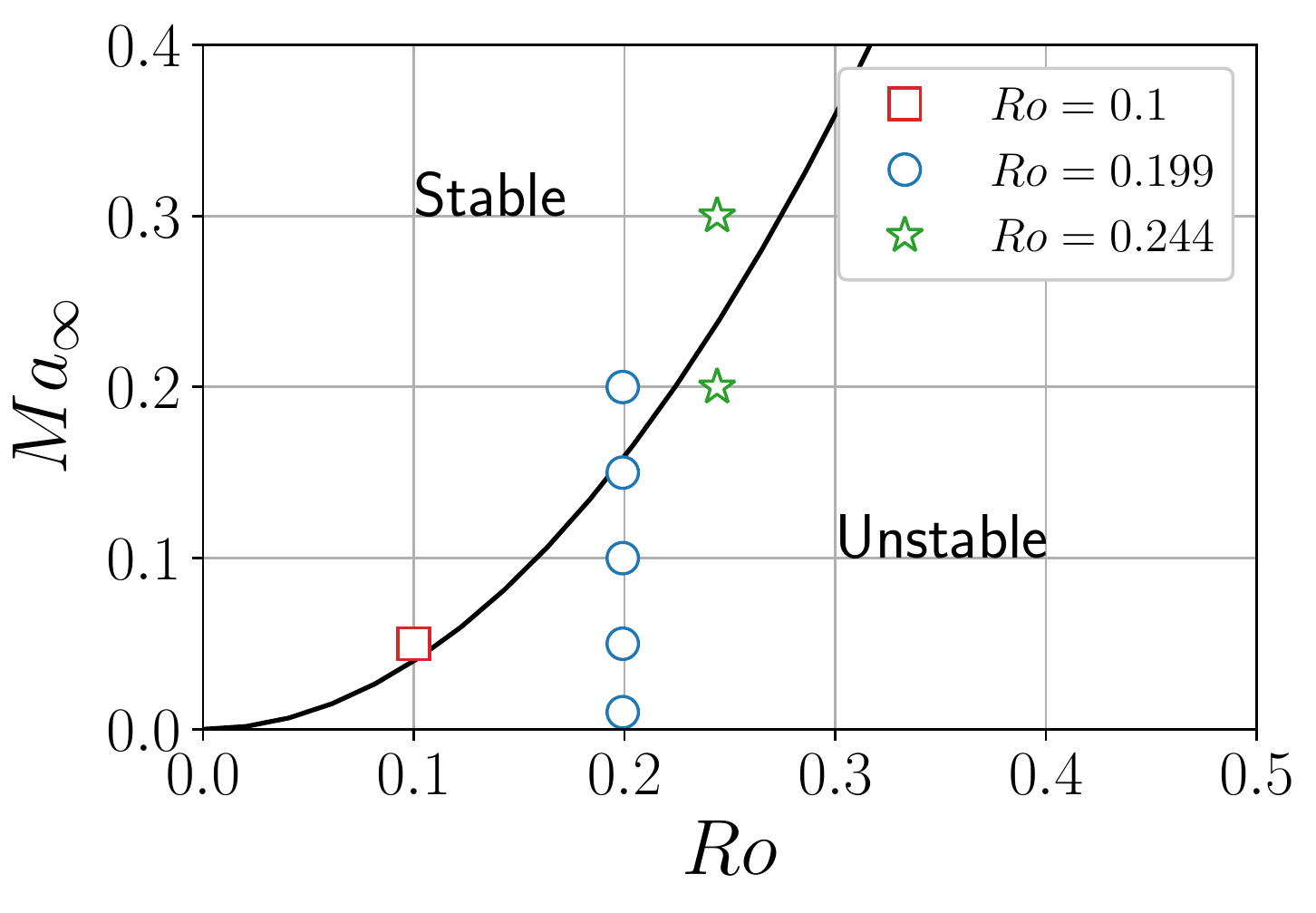}
\vspace*{-0.15in}
\caption{Stability curve as given in equation \ref{stability} and showing the conditions at different $Ro$ numbers. The black curve is the stability criterion defined by equation \ref{T_eq_ch4}}
\label{fig:stability}
\end{figure}

To study the effect of compressibility on the growth of the flow instability, the $Ma$ number is varied at a constant value of $Ro=0.199$ for both the $Ra$ numbers. Figure \ref{fig:Ma1e05} shows the time evolution of the most unstable mode for $Ra=10^5$ and $10^7$ cases at different $Ma$ numbers. The growth rate of the most unstable mode reduces with the increase in the value of $Ma$ number, and the flow becomes linearly stable at $Ma=0.2$ for both the $Ra$ numbers. The growth rate of the most unstable mode in the linear regime is further calculated at each $Ma$ number and plotted for both the $Ra$ numbers in figure \ref{fig:lambda}. At $Ma$ number of $0.2$, the growth rate of all the modes becomes negative for both the $Ra$ numbers. Thus, the compressibility effects lead to the suppression of instability which normally leads to the formation of convection rollers in the domain.
To further understand the stabilizing effect of $Ma$ number, a stability criterion for the onset of convection is derived. The flow stability is considered to be analogous to the static stability condition of the atmosphere \cite{cushman2011introduction}.
For the convection instability to grow, the temperature gradient in the system has to be larger than the adiabatic temperature gradient. Thus, for a stable system, the condition
\begin{equation}
\frac{dT}{dr} < \bigg(\frac{dT}{dr}\bigg)_{ad}
\label{T_eq_ch4}
\end{equation}
has to be satisfied. 
\begin{figure}[ht]
\centering
\includegraphics[width=0.85\textwidth]{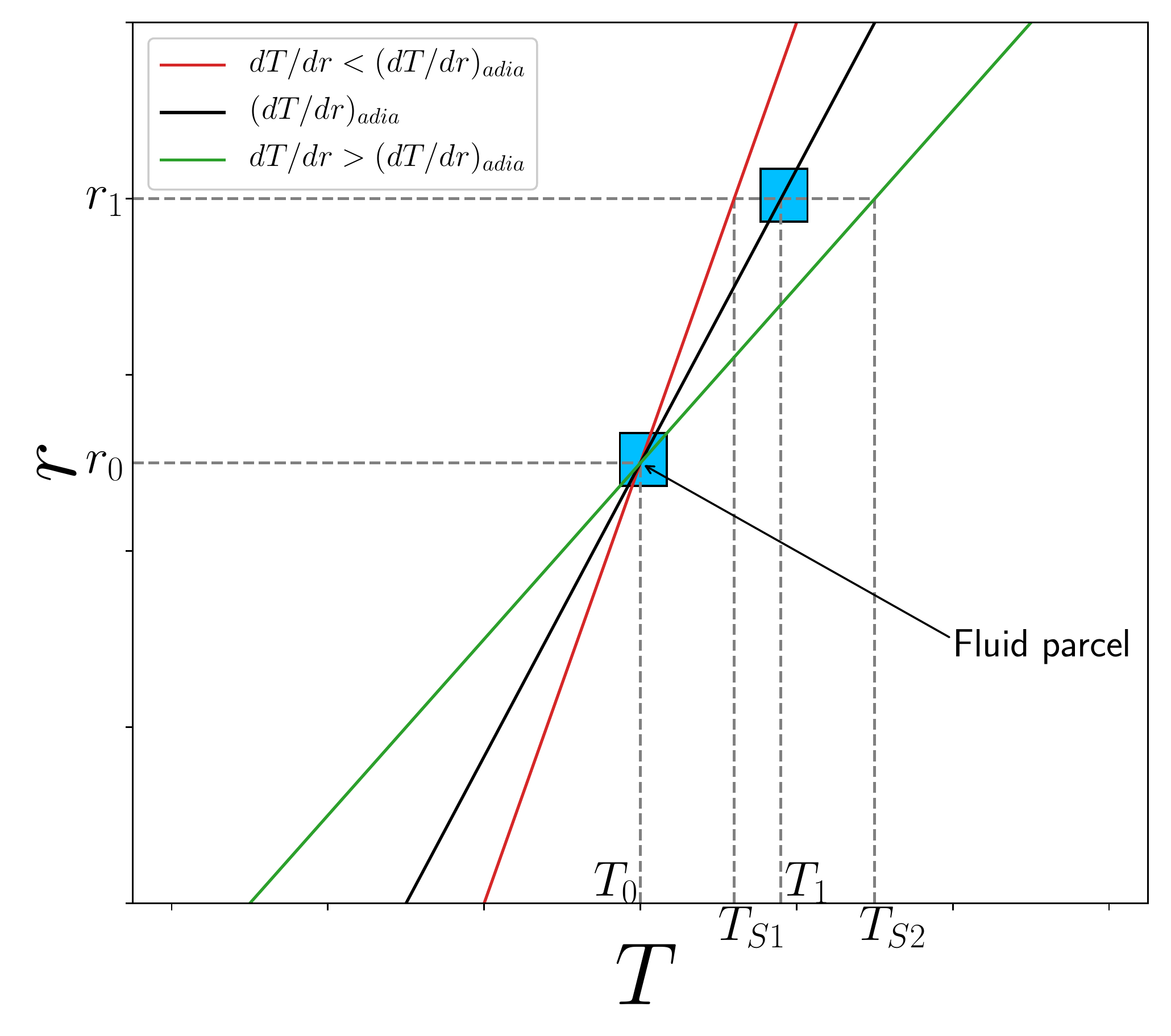}
\vspace*{-0.15in}
\caption{A schematic to illustrate the stability condition given in equation \ref{T_eq_ch4}, assuming the fluid parcel follows the adiabatic temperature profile. The temperature profiles shown here for different possible scenarios are just the representation to explain the condition given in equation \ref{T_eq_ch4}. The actual profiles might have different slopes.}
\label{fig:stab_deriv}
\end{figure}
To elaborate on the stability condition given in equation \ref{T_eq_ch4}, consider the upward and the downward motion of fluid parcels follows an adiabatic process, as shown in figure \ref{fig:stab_deriv}. It implies that the temperature of the fluid parcel can increase or decrease only by the work done during this motion and not by any heat transfer from the surroundings. To illustrate this, in figure \ref{fig:stab_deriv}, let us assume the initial position of the fluid parcel is at $r=r_0$ at temperature $T_0$. Due to centrifugal forces, the fluid parcel moves to a new position at $r=r_1$ and attains the temperature value of $T_1$ by following the adiabatic temperature profile. If the temperature gradient $(dT/dr < (dT/dr)_{adia})$ is less than the adiabatic one, then the upward motion of the fluid parcel (adiabatically) leads to an increase in its temperature (decrease in the density) compared to its surroundings, i.e $T_1 > T_{S1}$ as shown in figure \ref{fig:stab_deriv}. In the centrifugal force field, a low-density fluid parcel in the high-density surroundings is pushed back to its original state, and thus the system is stabilized. On the other hand, a temperature gradient greater than the adiabatic temperature gradient $(dT/dr > (dT/dr)_{adia})$ leads to an unstable situation as in this case the fluid parcel will have low temperature value (or high density) compared to the surroundings $(T_1 < T_{S2})$. In this scenario, the higher centrifugal force on the heavier fluid parcel allows the instability to grow, making the flow unstable. 
The stability criterion given in equation \ref{T_eq_ch4} can also be expressed in terms of non-dimensional parameters by using the first law of thermodynamics and the pressure balance equation with the centrifugal force (refer to Appendix \ref{App_A} for more details) in the system as

\begin{equation}
Ro < Ma_{\infty}^{1/2} \frac{(\gamma-1)^{1/4}}{2^{3/2}} \bigg(\frac{1+r_i/r_o}{1-r_i/r_o}\bigg)^{1/2}.
\label{stability}
\end{equation}
Figure \ref{fig:stability} shows the stability curve (black curve) as given by equation \ref{stability}. In the plane of $Ro$--$Ma_{\infty}$, the stability curve takes the parabolic shape ($Ma_{\infty} \approx Ro^2$) which separates the stable region from the unstable region. For both the $Ra$ numbers of $10^5$ and $10^7$, the conditions at $Ma=0.2$ and $Ro=0.199$ satisfy the above-mentioned stability criterion as shown in figure \ref{fig:stability}. 
To further test the validity of equation \ref{stability}, additional linear stability calculations were conducted at $Ra=10^7$ with $Ro$ numbers of $0.1$ and $0.244$. At the lower $Ro$ number of $0.1$, the flow is stable even at the low $Ma$ number of $0.05$ as it satisfies the condition of stability (equation \ref{stability}). At the higher $Ro$ number of $0.244$, the flow is observed to be unstable at $Ma=0.2$, with the wavenumber 16 being the most unstable. However, for this combination of $Ro$ and $Ra$, the flow becomes stable at a higher $Ma$ number of $0.3$, where it satisfies the stability criterion given in equation \ref{stability}. Thus, with the increase in the value of $Ro$ number, the value of $Ma$ number for the flow stability also increases.


\section{Conclusions}
In this study, we conducted linear stability analyses of centrifugal buoyancy-induced flow in a closed rotating cavity at $Ra=10^5$ and $10^7$. The propagation of instability was observed to be asymmetric in the azimuthal direction. The results of linear stability analysis showed the amplification of multiple modes in the linear regime, and the superposition of all these modes appeared as a localized disturbance. Further, the linearly most amplified modes obtained from LSA have much higher wavenumbers compared to the mode that appeared to dominate in the saturation state of DNS.
An increase in the $Ma$ numbers appeared to have a stabilizing effect and suppressed the formation of convection rollers. A stability criterion was proposed that showed that both Rossby and Mach number affect the stability of the rotating cavities. The flow Mach number imposed a constraint on the critical value of the Rossby number for the flow to be linearly stable.

\section{Acknowledgements}
This work was supported by resources provided by the Pawsey Supercomputing
Centre with funding from the Australian Government and the Government of
Western Australia.


\section{Appendix}
\subsection{Derivation for stability criteria}
\label{App_A}
As discussed in section \ref{eff_comp}, for the convection instability to grow, the temperature gradient in the system has to be larger than the adiabatic temperature gradient. Thus, for a stable system the condition
\begin{equation}
\frac{dT}{dr} < \bigg(\frac{dT}{dr}\bigg)_{ad}
\label{T_eq}
\end{equation}
has to be satisfied. 

To further derive the stability criteria, we consider the first law of thermodynamics which states that
\begin{equation}
\label{therm1}
dQ = du + dW = c_v dT + pdv .
\end{equation}
We can write $pdv = p d(1/\rho) = (-p/\rho^2)d\rho$ and further by using the equation of state $p=\rho R T$ and then $dp = \rho R dT + RT d\rho$. The final expression for $p dv$ will be

\begin{equation}
p dv = -\frac{p}{\rho^2} d\rho = -\frac{dp}{\rho} + R dT.
\end{equation}

Now we can rewrite the equation \ref{therm1} as
\begin{equation}
\label{eq2}
dQ = (c_p - R)dT -\frac{dp}{\rho} + R dT.
\end{equation}

Since we are assuming the process to be adiabatic, thus $dQ=0$ and equation \ref{eq2} becomes
\begin{equation}
\label{eq3}
\frac{dp}{\rho} = c_p dT.
\end{equation}
Further, the equation of hydrostatic balance in the rotating flows can be written as
\begin{equation}
\label{eq4}
\frac{dp}{dr} = \rho \Omega^2 r.
\end{equation}

By using the equations \ref{eq3} and \ref{eq4}, the adiabatic temperature gradient is equal to

\begin{equation}
\label{eq5}
\bigg(\frac{dT}{dr} \bigg)_{adia} = \frac{\Omega^2 r}{c_p}.
\end{equation}

Thus, by using equation \ref{eq5}, the expression for stability criteria becomes 
\begin{equation}
\frac{dT}{dr} < \bigg(\frac{\Omega^2 r}{c_p}\bigg).
\end{equation}
Integrate equation \ref{eq5} from the inner radius to the outer radius (refer to figure \ref{fig:sch}).

\begin{equation}
\begin{aligned}
&\ \int_{T_C}^{T_H} dT < \int_{r_i}^{r_o} \bigg(\frac{\Omega^2 r}{c_p}\bigg) dr \\
&\ \Delta{T} = T_H - T_C < \frac{\Omega^2}{2 c_p} (r_o^2 - r_i^2). \\
\end{aligned}
\label{eq6}
\end{equation}
A stability criterion similar to equation \ref{eq6} was also proposed by Kilfoil and Chew \cite{kilfoil2009modelling}, and they used it to improve their turbulence model. However, in this study, we have provided a physical interpretation and confirmation about the validity of the stability criterion from linear stability results in section \ref{eff_comp}.
Further, by using the definition of Mach number and Rossby number given in equations \ref{ND} and \ref{Ro_eq}, we can write the \ref{eq6} as
\begin{equation}
Ro < Ma_{\infty}^{1/2} \frac{(\gamma-1)^{1/4}}{2^{3/2}} \bigg(\frac{1+r_i/r_o}{1-r_i/r_o}\bigg)^{1/2}.
\end{equation}

\section*{References}

\bibliography{myrefs}

\end{document}